\documentclass[a4paper,11pt]{article}
\pdfoutput=1

\usepackage{jcappubmod}

\usepackage[T1]{fontenc}
\usepackage{nicefrac}

\title{Late-time acceleration by a residual cosmological constant from sequestering vacuum energy in ultimate collapsed structures}

\author[a]{Lucas Lombriser}

\affiliation[a]{D\'{e}partement de Physique Th\'{e}orique, Universit\'{e} de Gen\`{e}ve, \\ 24 quai Ernest Ansermet, 1211 Gen\`{e}ve 4, Switzerland}

\emailAdd{lucas.lombriser@unige.ch}

\abstract{
The sequestering mechanism has been proposed as a remedy to the \emph{old} cosmological constant problem of the non-gravitating vacuum energy in the matter sector.
Here it is shown that an extension of this global constraint mechanism arises naturally from an averaging condition for residual cosmological constants produced in different patches of the Universe.
The new mechanism naturally yields the small observed value that gives rise to cosmic acceleration at late times, hence, addressing the \emph{new} cosmological constant problem.
Hereby the halo model picture is adopted with all matter content ultimately residing in the last collapsed structures formed.
Sequestering in these collapsing patches produces the desired average residual, where a uniform prior on our present location in the collapse predicts $\Omega_{\Lambda}=0.697$.
Finally, a fluctuation of the local residual from the cosmological average can naturally give rise to a locally enhanced Hubble constant.
}

\begin{document}
\maketitle
\flushbottom

\section{Introduction} \label{sec:intro}

The physical nature underlying the observed late-time accelerated expansion of our Universe~\cite{Perlmutter:1998np,Riess:2016jrr} remains a difficult puzzle to cosmology.
Generally the effect is attributed to a cosmological constant $\Lambda$, which is thought to arise from vacuum fluctuations.
Quantum theoretical calculations of this vacuum energy contribution, however, deviate from the observed value by $\gtrsim50$ orders of magnitude~\cite{Weinberg:1988cp,Martin:2012bt}.
Alternatively, cosmic acceleration has been conjectured to be due to a dark energy field permeating the Universe or a breakdown of general relativity at large scales~\cite{Clifton:2011jh,Koyama:2015vza,Joyce:2016vqv}.
The dynamics of dark energy however must be fine-tuned to closely mimic a cosmological constant~\cite{Ade:2015xua}, and the equality between the speeds of gravity and light confirmed by the LIGO/Virgo GW170817 measurement~\cite{TheLIGOScientific:2017qsa,Monitor:2017mdv,GBM:2017lvd} combined with observations of the large-scale structure poses hard challenges to the concept of cosmic self acceleration from a genuine modification of gravity~\cite{Lombriser:2015sxa,Lombriser:2016yzn}.
While these models have been invoked to address the enigma of cosmic acceleration, and sometimes the \emph{Why Now?} conundrum of the presently coincident energy densities of the cosmological constant and the matter content~\cite{Martin:2012bt,Lombriser:2017cjy}, both attributed to the \emph{new} cosmological constant problem, they usually do not provide a mechanism to circumvent the \emph{old} cosmological constant problem of the non-gravitating vacuum energy  (however, see, e.g., Ref.~\cite{Appleby:2018yci}).

A mechanism for the sequestering of the vacuum energy from the matter sector has been proposed as a remedy to the old cosmological constant problem in Ref.~\cite{Kaloper:2013zca}.
It employs a global constraint on the residual cosmological constant affecting gravitational dynamics from a historical spacetime average of the matter content in the Universe that turns it radiatively stable.
A local version of the sequestering mechanism has also been developed in Ref.~\cite{Kaloper:2015jra}.
While the mechanism provides a promising evasion of vacuum energy contributions to gravitational dynamics, it does not predict the required collapse of the Universe by itself, and in the global version the residual cosmological constant produced by a future collapse is generally too small to explain the observed current cosmic acceleration.
In contrast, in the local version a constant correction term to this small residual can be set arbitrarily by measurement.

This paper proposes an extension of the global sequestering mechanism to account for patchy matter distributions in the Cosmos that only become homogeneous on large scales and that undergo gravitational collapse.
Thereby a halo model view is adopted that assigns all matter in the Universe to virialised halos of different sizes.
In the future, structure formation through collapse ceases, which allows to assign the massive patches to ultimate collapsed structures.
The residual cosmological constants in these structures determine the averaged cosmological constant in the background expansion, which is computed here and compared against observations.

The paper is organised as follows.
Sec.~\ref{sec:sequestering} briefly reviews the dynamical aspects of the sequestering mechanism of Ref.~\cite{Kaloper:2013zca}.
The extended mechanism for collapsing patches of the overall matter distribution is developed in Sec.~\ref{sec:extendedseq}.
The nonlinear evolution of these patches is discussed in Sec.~\ref{sec:sphcoll} and results for the residual cosmological constant produced by the mechanism are presented in Sec.~\ref{sec:residualcc}.
Sec.~\ref{sec:cosmicacc} discusses expectations for the relative size of the energy densities of the observed cosmological constant and the matter content, considering aspects of the star-formation history or our likely location in the formation of an ultimate collapsed structure.
It also provides a brief discussion of further observational implications.
Finally, conclusions are drawn in Sec.~\ref{sec:conclusions}.

\section{Sequestering Standard Model vacuum energy} \label{sec:sequestering}

The Einstein-Hilbert action for the sequestering of vacuum energy from the matter sector in a minimal alteration of standard general relativity is given by~\cite{Kaloper:2013zca}
\begin{equation}
 S = \int d^4x \sqrt{-g} \left[ \frac{M_{\rm Pl}^2}{2} R - M_{\rm Pl}^2 \Lambda + \lambda^4 \mathcal{L}_{\rm m}(\lambda^{-2}g^{\mu\nu},\Phi) \right] + \sigma\left( \frac{M_{\rm Pl}^2\Lambda}{\lambda^4\mu^4} \right)
\end{equation}
with units $\hbar=c=1$, where $\sigma$ is the global sequestering term.
$\Lambda$ is an arbitrary classical cosmological constant and the constant parameter $\lambda$ controls the hierarchy between the matter and Planck scales.
These are treated as global variables to be included in the variational principle whereas the scale $\mu$ is determined phenomenologically.
Variations of the action with respect to $\Lambda$ and $\lambda$ yield
\begin{eqnarray}
 \frac{\sigma'}{\lambda^4\mu^4} & = & \int d^4x\sqrt{-g} \,, \\
 4 M_{\rm Pl}^2 \Lambda \frac{\sigma'}{\lambda^4\mu^4} & = & \int d^4x\sqrt{-g} T \,,
\end{eqnarray}
where $T \equiv T_{\ \mu}^{\mu}$, $\tilde{T}_{\mu\nu} = -2 [\delta(\sqrt{-\tilde{g}}\mathcal{L}_{\rm m})/\delta\tilde{g}^{\mu\nu}]/\sqrt{-\tilde{g}}$ with $\tilde{g}^{\mu\nu}=\lambda^{-2}g^{\mu\nu}$, and $T^{\mu}_{\ \nu}=\lambda^4\tilde{T}^{\mu}_{\ \nu}$.
From the combination of the two equations one obtains
\begin{equation}
 \Lambda = \frac{1}{4 M_{\rm Pl}^2} \frac{\int d^4x \sqrt{-g} T}{\int d^4x \sqrt{-g}} \equiv \frac{1}{4 M_{\rm Pl}^2} \left\langle T \right\rangle \,. \label{eq:sequestering}
\end{equation}
Finally, variation of the action with respect to $g_{\mu\nu}$ gives
\begin{equation}
 R^{\mu}_{\ \nu} - \frac{1}{2}R \delta^{\mu}_{\ \nu} + \frac{1}{4 M_{\rm Pl}^2} \langle T \rangle \delta^{\mu}_{\ \nu} = \frac{1}{M_{\rm Pl}^2} T^{\mu}_{\ \nu} \,. \label{eq:fieldeqs}
\end{equation}
It can now be observed that the contribution of a bare cosmological constant $\Lambda_{\rm bare}$ or a quantum vacuum correction $\Lambda_{\rm vac}$ at any loop order in the matter sector cancels out in these field equations.
Hence, the vacuum energy does not gravitate in the sequestering model, which may provide a remedy to the old cosmological constant problem.
This can be seen explicitly by replacing $\mathcal{L}_{\rm m}\rightarrow\mathcal{L}_{\rm m} + \Lambda_{\rm bare} + \Lambda_{\rm vac}$.
Since $\langle \Lambda_{\rm bare} + \Lambda_{\rm vac} \rangle = \Lambda_{\rm bare} + \Lambda_{\rm vac}$, its contribution to the left- and right-hand sides of Eq.~(\ref{eq:fieldeqs}) cancel.
Further particulars regarding the sequestering of the vacuum energy can be found in Refs.~\cite{Kaloper:2013zca,Kaloper:2014dqa}.
Note that there is also a local version of the vacuum energy sequestering mechanism~\cite{Kaloper:2015jra},
where a constant correction term of arbitrary value contributes to Eq.~(\ref{eq:fieldeqs}) that can be fixed to match observations.
Rather than on the quantum aspects of the mechanism, this paper shall instead focus on the generation of the radiatively stable $\Lambda$ from the dynamics of $T$ in the global version.

Importantly, the four-volume term in Eq.~(\ref{eq:sequestering}) grows large for an old universe.
Should one wish to match the residual with the observed cosmological constant and assuming $\Lambda$CDM with Planck cosmological parameters~\cite{Ade:2015xua}, as opposed to the local mechanism, one finds that the Universe should have undergone an immediate collapse at the scale factor $a=0.712$, corresponding to an age of $0.65H_0^{-1}$, hence, about 4.4~Gyr in the past~\cite{Lombriser:2018}.
In contrast, an immediate collapse today would account for 40\% of the observed cosmological constant.
While it is interesting that this value is in $\mathcal{O}(1)$ agreement with the observed value, standard cosmology does not foresee an imminent collapse of our Universe.
Therefore the global version of the sequestering mechanism requires some additional fields or an extension to naturally produce the required collapse in Eq.~(\ref{eq:sequestering}) and bring the residual cosmological constant into agreement with observations.
While this can be achieved by designing an exotic dark energy or modified gravity model with the correct fluid properties~\cite{Tsukamoto:2017brj,Lombriser:2018}, Sec.~\ref{sec:main} will present an extended global sequestering mechanism suitable to patchy matter distributions that provides the desired properties.

\section{Late-time acceleration by the residual cosmological constant from ultimate collapsed structures} \label{sec:main}

Let us now develop an extension to the sequestering mechanism of Ref.~\cite{Kaloper:2013zca} to render it suitable for collapsing patches in an overall matter distribution such that the residual cosmological constants produced in their evolution can be computed and compared against observations.
The extended mechanism is introduced in Sec.~\ref{sec:extendedseq}, the nonlinear evolution of the collapsing patches is discussed in Sec.~\ref{sec:sphcoll}, and results for the residual cosmological constants produced by the mechanism are presented in Sec.~\ref{sec:residualcc}.
Finally, in Sec.~\ref{sec:cosmicacc} the relative size of the energy densities of the observed cosmological constant and the matter content are compared under consideration of the star-formation history or the likelihood of our location in the formation of an ultimate collapsed structure.
For simplicity, only the matter components and cosmological constant shall be considered in this analysis, where it has been checked that the introduction of a radiation component only marginally affects the results presented in Secs.~\ref{sec:residualcc} and \ref{sec:cosmicacc}.
Furthermore, the cosmological background shall be assumed to be spatially flat.

\subsection{Global constraints from ultimate collapsed structures} \label{sec:extendedseq}

We shall adopt the halo model view for the matter content in the Universe, in that all matter is assumed to reside in virialised halos.
This will enable the construction of a sequestering mechanism for patchy matter distributions.
As discussed in Sec.~\ref{sec:sphcoll}, the formation of virialised halos through spherical collapse will come to an end in the future when the expansion rate of the cosmological background becomes too large to be counteracted by the self gravity of an overdense region, at which time the ultimate collapsed structures will be formed.
The matter in the Universe at a given time $\mathcal{M}^{(3)}$ can be divided into patches $\mathcal{U}_i^{(3)}$ that will end up in a given halo at the end of structure formation.
In the early universe, at some $t_{\rm ini} \ll t_0$, the matter distribution was nearly perfectly homogeneous, and therefore these patches approximately covered the entire space,
\begin{equation}
 \mathcal{M}_{\rm ini}^{(3)} \simeq \bigcup_i \mathcal{U}_{i,{\rm ini}}^{(3)} \,.
\end{equation}
For the average cosmological constant over these patches, this implies
\begin{equation}
 \Lambda \simeq \frac{ \sum_i \Lambda_i\int_{\mathcal{U}_{i,{\rm ini}}^{(3)}} dV_3 }{ \int_{\mathcal{M}_{{\rm ini}}^{(3)}} dV_3 } \simeq \frac{ \sum_i \Lambda_i\int_{\mathcal{U}_{i,{\rm ini}}^{(3)}} dV_3 }{ \sum_i \int_{\mathcal{U}_{{\rm ini},i}^{(3)}} dV_3 } \,, \label{eq:initial}
\end{equation}
where $\Lambda_i$ denotes the cosmological constant on $\mathcal{U}_i$.

Due to the collapse of matter into halos, at a later time, the massive spaces $\mathcal{U}_i^{(3)}$ do not fill $\mathcal{M}^{(3)}$ anymore, leaving behind empty regions.
The cosmological constant, however, should not be confined to $\mathcal{U}_i^{(3)}$ and extend to these empty patches.
We shall therefore require that the cosmological constant $\Lambda$ in the total patch left behind by the collapsing structures should match the sum of the individual volume-averaged cosmological constants $\Lambda_i$ in the regions left behind by $\mathcal{U}_i$.
For this purpose, let us define the four-dimensional subspaces $\mathcal{M}_i$ that match $\mathcal{U}_i$ at early times, $\mathcal{M}_i\equiv\left\{ x^{\mu}\in\mathcal{M} | \bigcup_i\mathcal{M}_i=\mathcal{M} \land \mathcal{M}_{i,{\rm ini}}^{(3)} \simeq  \mathcal{U}_{i,{\rm ini}}^{(3)} \right\}$.
The condition on $\Lambda$ can then be expressed as
\begin{equation}
 \Lambda\simeq\frac{ \sum_i \int_{\mathcal{M}_i\backslash\mathcal{U}_i} dV_4 \Lambda_i }{ \int_{\mathcal{M}\backslash\bigcup_i\mathcal{U}_i} dV_4 } = \frac{ \sum_i \Lambda_i \int_{\mathcal{M}_i\backslash\mathcal{U}_i} dV_4 }{ \sum_i \int_{\mathcal{M}_i\backslash\mathcal{U}_i} dV_4} = \frac{\sum_i \Lambda_i \bar{\beta}_i}{\sum_i \bar{\beta}_i} = \frac{\sum_i \Lambda_i \bar{\beta}_i}{\bar{\beta}} \,, \label{eq:condition}
\end{equation}
where we have defined $\bar{\beta}_i\equiv\int_{\mathcal{M}_i\backslash\mathcal{U}_i} dV_4$ and $\bar{\beta}\equiv\int_{\mathcal{M}\backslash\bigcup_i\mathcal{U}_i} dV_4$.
Furthermore, defining
$\beta_i\equiv\int_{\mathcal{U}_i} dV_4$ and $\alpha_i\equiv\int_{\mathcal{M}_i} dV_4$ with $\beta$ and $\alpha$ denoting the respective sums, it holds that $\alpha_i=\beta_i+\bar{\beta}_i$ and $\alpha=\beta+\bar{\beta}$, and one finds the equivalent condition
\begin{equation}
 \Lambda \simeq \frac{\sum_i \Lambda_i \beta_i}{\beta} = \frac{\sum_i \Lambda_i \beta_i}{\sum_i \beta_i} = \frac{ \sum_i \Lambda_i \int_{\mathcal{U}_i} dV_4 }{ \sum_i \int_{\mathcal{U}_i} dV_4} = \frac{ \sum_i \int_{\mathcal{U}_i} dV_4 \Lambda_i }{ \int_{\bigcup_i\mathcal{U}_i} dV_4 } \label{eq:condition2}
\end{equation}
when requiring $\Lambda = \alpha^{-1}\sum_i\alpha_i\Lambda_i$.

From these considerations, the sequestering mechanism discussed in Sec.~\ref{sec:sequestering} can now be extended to patchy matter distributions,
\begin{eqnarray}
 S & = & \int_{\mathcal{M}\backslash\bigcup_i\mathcal{U}_i} dV_4 \left[ \frac{R}{2} - \Lambda - \Lambda_{\rm seq} + \lambda_{\rm seq}^4 \mathcal{L}_{\rm m}(\lambda_{\rm seq}^{-2}g^{\mu\nu},\Phi) \right] \nonumber\\
  & & + \sum_i \left\{ \int_{\mathcal{U}_i} dV_4 \left( \frac{R}{2} - \Lambda_i + \lambda_i^4 \mathcal{L}_{\rm m}(\lambda_i^{-2}g^{\mu\nu},\Phi) \right] \right\} + b.t. \nonumber\\
 & & + \lambda_{\rm B} \left( \frac{\Lambda}{\lambda^4} - \sum_i \frac{\Lambda_i}{\lambda_i^4} \right)  + \sigma\left(\frac{\Lambda_{\rm seq}}{\lambda_{\rm seq}^4}\right) \,,
\end{eqnarray}
where we have set $M_{\rm Pl}=1$ and $\mu=1$ for convenience.
The parameter $\lambda_B$ corresponds to a Lagrange multiplier and will ensure that the condition~(\ref{eq:condition2}) is satisfied, which naturally gives rise to a sequestering term.\footnote{Note that one may also define $\sigma(\lambda_{\rm seq}^{-4}\Lambda_{\rm seq}) \equiv \lambda_{\rm B} \lambda_{\rm seq}^{-4}\Lambda_{\rm seq}$, in which case the global sequestering term can be attributed to the Lagrange multiplier constraint.
This, however, requires a large and long-living matter volume (with large $\lambda^{-4}=-\lambda_{\rm B}^{-1}\beta$) to suppress the contribution $-\lambda^4(\Lambda_{\rm bare}+\Lambda_{\rm vac})$ to $\Lambda$ governing the dynamics in the empty patch and to recover condition~(\ref{eq:condition2}).}
The action is varied in the $g_{\mu\nu}$ of the different patches as well as in $\Lambda_{\rm seq}$, $\lambda_{\rm seq}$, $\Lambda_i$, $\lambda_i$, and $\lambda_{\rm B}$.
From variations of the metric in $\mathcal{M}\backslash\bigcup_i\mathcal{U}_i$ and of $\Lambda_{\rm seq}$ and $\lambda_{\rm seq}$, one finds again that contributions of $\Lambda_{\rm bare}$ and $\Lambda_{\rm vac}$ cancel in the field equations, and since this space is otherwise empty, the residual cosmological constant is $\Lambda$, which is given by the variation in $\lambda_{\rm B}$ in terms of $\Lambda_i$, $\lambda_i$, and $\lambda$.
Variations in the metric on $\mathcal{U}_i$ and in $\Lambda_i$ and $\lambda_i$ in contrast yield $\Lambda_i=\langle T \rangle_i$/4, which involves the computation of the fully nonlinear $\langle T \rangle_i$ (Secs.~\ref{sec:sphcoll} and \ref{sec:residualcc}), and contributions of $\Lambda_{\rm bare}$ and $\Lambda_{\rm vac}$ again cancel in the field equations.
Furthermore, the variation in $\Lambda_i$ yields $\lambda_i^{-4}=-\lambda_{\rm B}^{-1}\beta_i$, and together with the variation in $\lambda_{\rm B}$ one finds that the average cosmological constant in the Universe $\alpha^{-1}\sum_i\alpha_i\Lambda_i$ is given by
$(\bar{\beta} + C\beta)\Lambda/(\bar{\beta}+\beta)$ with constant $C\equiv-\lambda_{\rm B}\lambda^{-4}\beta^{-1}$.
Given the initial condition~(\ref{eq:initial}), the physically most relevant choice is $C=1$.
More strictly, since the matter patches collapse in the future, the average cosmological constant must match that of the empty patch $\Lambda$, which implies $C=1$ and $\lambda^{-4}=-\lambda_{\rm B}^{-1}\beta = -\lambda_{\rm B}^{-1}\sum_i\beta_i = \sum_i\lambda_i^{-4}$.
Hence, the Lagrange multiplier recovers condition~(\ref{eq:condition2}), which will be used in Sec.~(\ref{sec:residualcc}) to compute the residual cosmological constant, and with that also condition~(\ref{eq:condition}).
Finally, boundary terms have also been added to the action to ensure a well-defined variational principle for the different $g_{\mu\nu}$. These are just the standard Gibbons-Hawking-York terms for the different manifolds with boundary, producing the usual field equations of Sec.~\ref{sec:sequestering} and introduced in local sequestering in Ref.~\cite{Kaloper:2016yfa} (also see Ref.~\cite{BeltranJimenez:2017tkd} on how boundary terms can be dispensed with by a geometrical reformulation of general relativity).

\subsection{Spherical collapse} \label{sec:sphcoll}

The formation of nonlinear cosmological structures can be described with the spherical collapse model, which approximates forming halos by spherically symmetric top-hat overdensities.
Following energy-momentum conservation, $\nabla^{\mu}T_{\mu\nu}=0$, these are then evolved according to the nonlinear continuity and Euler equations from an initial era to the time of their collapse.
In comoving spatial coordinates and for a pressureless non-relativistic matter fluid the continuity and Euler equations are given by~\cite{peebles:80}
\begin{eqnarray}
 \dot{\delta} + \frac{1}{a}\nabla\cdot(1+\delta)\mathbf{v} & = & 0 \,, \\
 \dot{\mathbf{v}} + \frac{1}{a}\left(\mathbf{v}\cdot\nabla\right)\mathbf{v} + H \mathbf{v} & = & -\frac{1}{a}\nabla\Psi \,,
\end{eqnarray}
where dots indicate derivatives with respect to physical time, $\delta\equiv\delta \rho_{\rm m}/\bar{\rho}_{\rm m}$, $\Psi\equiv\delta g_{00}/(2g_{00})$ denotes the gravitational potential, and $H\equiv\dot{a}/a$ is the Hubble parameter.
In combination, the two equations yield
\begin{equation}
 \ddot{\delta} + 2H\dot{\delta} - \frac{1}{a^2}\nabla_i\nabla_j(1+\delta)v^iv^j = \frac{1}{a^2}\nabla_i(1+\delta)\nabla^i\Psi \,.
\end{equation}
Adopting the spherical top-hat approximation, the velocity simply becomes $\mathbf{v}=A(t)\mathbf{r}$ and
\begin{equation}
 \frac{1}{a^2}\nabla_i\nabla_jv^iv^j=\frac{4}{3}\frac{\dot{\delta}^2}{(1+\delta)^2}
\end{equation}
follows from the continuity equation, which yields the spherical collapse equation
\begin{equation}
 \ddot{\delta} + 2H\dot{\delta} - \frac{4}{3}\frac{\dot{\delta}^2}{(1+\delta)} = \frac{1+\delta}{a^2}\nabla^2\Psi \,. \label{eq:sphcoll}
\end{equation}
Re-expressing the equation in terms of the physical top-hat radius at $a$, $\zeta(a)$, and employing mass conservation $M=(4\pi/3)\bar{\rho}_{\rm m}(1+\delta)\zeta^3= (4\pi/3)\rho_{\rm m}\zeta^3$, the evolution equation for the spherical shell at the edge of the top hat becomes
\begin{equation}
 \frac{\ddot{\zeta}}{\zeta} = H^2 + \dot{H} - \frac{1}{3a^2}\nabla^2\Psi = H^2 + \dot{H} -\frac{\kappa^2}{6} \delta\rho_{\rm m} \,, \label{eq:shellevol}
\end{equation}
where $ \rho_{\rm m} \equiv \bar{\rho}_{\rm m} + \delta\rho_{\rm m} \equiv \bar{\rho}_{\rm m} (1 + \delta)$ and from energy conservation $\bar{\rho}_{\rm m} = \bar{\rho}_{\rm m 0}a^{-3}$ with $\bar{\rho}_{\rm m 0} \equiv \bar{\rho}_{\rm m}(a=1)$.
Note that the background matter density is an average over all collapsing matter distributions with $\bar{\rho}_{\rm m} = \left( \sum_i \int_{\mathcal{U}_i^{(3)}} dV_3 \rho_{\rm m} \right) / \int_{\mathcal{M}^{(3)}}dV_3$, which follows from mass conservation, and the background cosmological constant is determined by condition~(\ref{eq:condition2}).
This therefore reproduces the usual Friedmann equations describing the evolution of the cosmological background in standard cosmology.

For a comoving top-hat radius $r_{\rm th}$ with $\zeta_i=a_i r_{\rm th}$ at an initial scale factor $a_i \ll 1$, mass conservation further implies $\bar{\rho}_{\rm m} a^3 r_{\rm th}^3 = \rho_{\rm m}\zeta^3$.
With the definition of a dimensionless physical top-hat radius $y\equiv\zeta/(a\:r_{\rm th})$, where $\rho_{\rm m}/\bar{\rho}_{\rm m} = y^{-3}$, Eq.~(\ref{eq:shellevol}) can be expressed as
\begin{equation}
 y'' + \left( 2 + \frac{H'}{H} \right) y' + \frac{1}{2} \Omega_{\rm m}(a) \left( y^{-3} - 1 \right) y = 0 \,, \label{eq:y}
\end{equation}
where primes denote derivatives with respect to $\ln a$ and $\Omega_{\rm m}(a) \equiv \kappa^2 \bar{\rho}_{\rm m}/(3H^2)$.
Eq.~(\ref{eq:y}) can be solved with initial conditions in the matter-dominated regime ($a_i=10^{-2}$), $y_i \equiv y(a_i) = 1 - \delta_i/3$ and $y_i' = - \delta_i/3$, which will be used in Sec.~\ref{sec:residualcc} to obtain the residual cosmological constant from the extended sequestering mechanism of Sec.~\ref{sec:extendedseq} in a patch $\mathcal{U}_i$.

Importantly, the collapse is a competition between the expansion of the cosmological background and the self gravity of the massive patch.
There is a minimal $\delta_i$ beyond which the expansion rate in the future will exceed the effect of self gravitation such that collapsed structures can no longer form.
Furthermore, note that no distinction will be made on whether the collapse implies the formation of a virialised halo or of a singularity, and we will simply refer to these last entities formed as ultimate collapsed structures.

\subsection{The residual cosmological constant} \label{sec:residualcc}

Having extended the sequestering mechanism to patchy matter distributions in Sec.~\ref{sec:extendedseq} and having derived the evolution equation for the collapsing regions $\mathcal{U}_i$, the residual cosmological constants from these regions can now be computed.
For simplicity, we shall first assume that the residual cosmological constants are equal among the different patches with perfect homogeneity in the matter distribution across the regions at some $t_{\rm ini}$.
Thus, $\Lambda_i=\Lambda\equiv\Lambda_{\rm res}$ $\forall i$.
Note, however, that in general this must not be the case and that, for instance, the cosmological constant in the local patch may differ from the averaged background value $\Lambda$ (Sec.~\ref{sec:extendedseq}), which may have interesting observational implications (Sec.~\ref{sec:observations}).

\begin{figure}
 \resizebox{0.5075\textwidth}{!}{
 \includegraphics{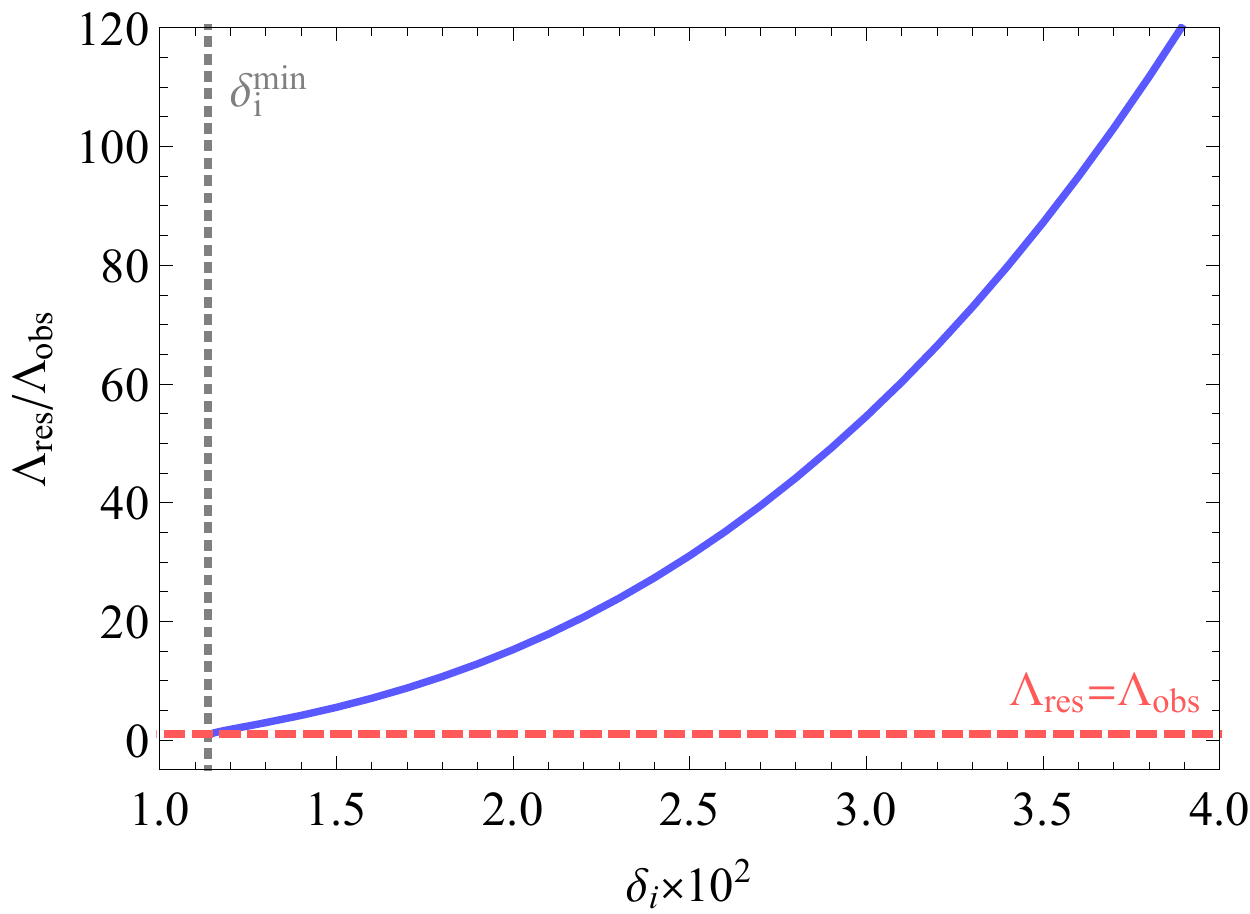}
 }
 \resizebox{0.4925\textwidth}{!}{
 \includegraphics{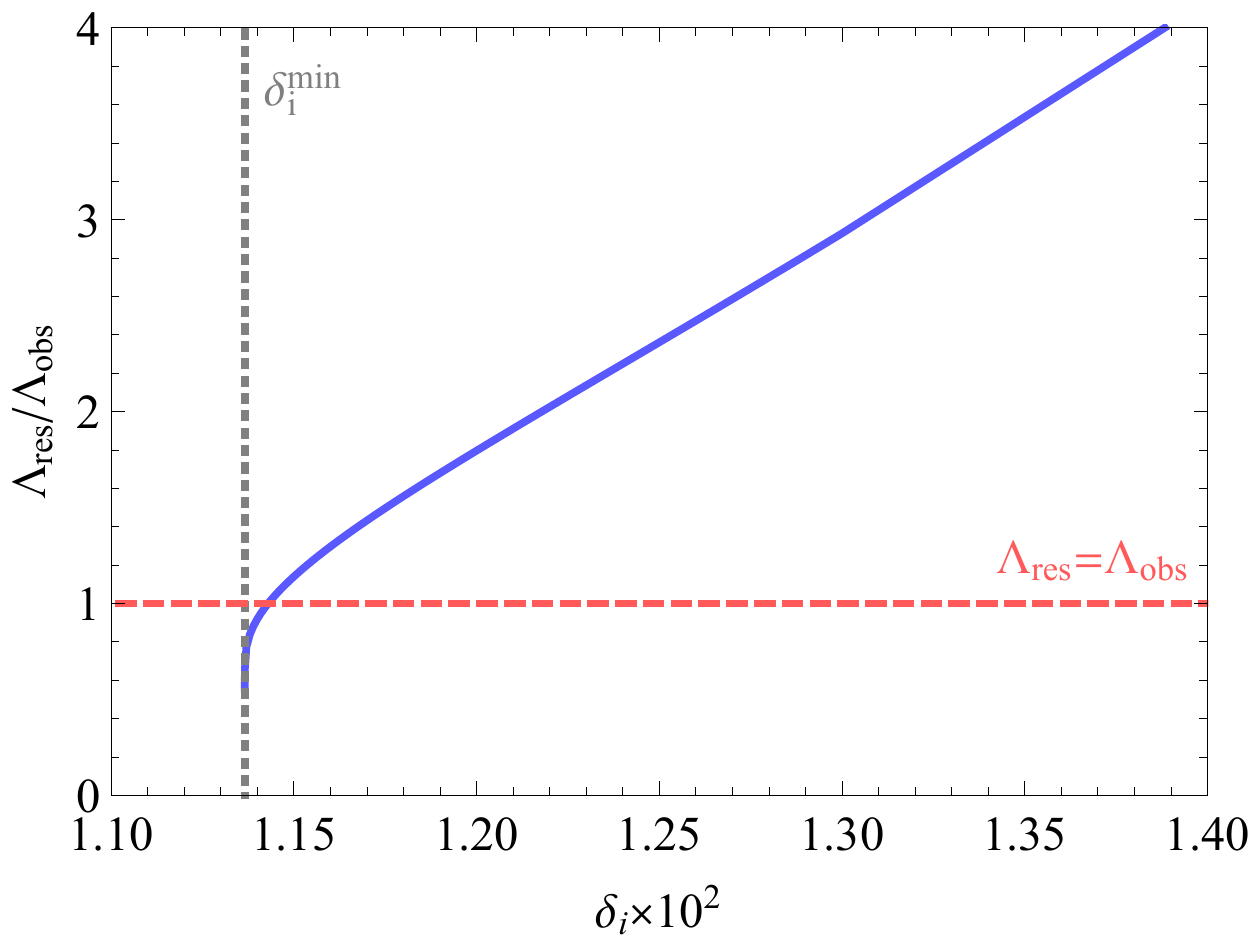}
 }
\caption{
The ratio $\Lambda_{\rm res}/\Lambda_{\rm obs}$ as a function of the initial top-hat overdensity $\delta_i$ at $a_i=10^{-2}$ with normalisation $a_{\rm eq}\equiv1$.
Larger values of $\delta_i$ collapse earlier in the history of the Universe.
The ultimate structures collapse at $\delta_i^{\rm min}$.
\emph{Left panel:} The residual cosmological constant from ultimate collapsed structures is naturally driven to the observed value.
\emph{Right panel:} A fine-tuning of $\delta_i$ may approximate $\Lambda_{\rm res}/\Lambda_{\rm obs}=\nicefrac{1}{2}$.
Small fluctuations around $\delta_i$ among the different collapsing patches are expected to cause small fluctuations of the observed cosmological constant in different patches of the sky.
}
\label{fig:ratio}
\end{figure}

To obtain $\Lambda_{\rm res}$, one needs to compute
\begin{eqnarray}
 \Lambda_{\rm res} = \left\langle T \right\rangle & = & \frac{1}{4} \frac{\int d^4x \sqrt{-g} T}{\int d^4x \sqrt{-g}} 
 = \frac{1}{4} \frac{\int dt \int d\tilde{\zeta} \: \tilde{\zeta}^2 \rho_{\rm m} }{\int dt \int d\tilde{\zeta} \: \tilde{\zeta}^2} \nonumber\\
 & = & \frac{1}{4} \frac{ \int dt \: \rho_{\rm m} \zeta^3}{\int dt \: \zeta^3}
 = \frac{\bar{\rho}_{\rm m0}}{4} \frac{\int d\ln a \: H^{-1}}{\int d\ln a \: H^{-1} a^3 y^3}
\end{eqnarray}
on the top hat.
Note that neither the evolution of $y$ nor the value of $\Lambda_{\rm res}$ depend on the comoving top-hat radius $r_{\rm th}$ and hence on the mass of the collapsed patch.
$\Lambda_{\rm res}$ however depends on the initial overdensity $\delta_i$.
Generally, $\delta_i$ should recover
\begin{equation}
 \frac{\Lambda_{\rm res}}{\Lambda_{\rm obs}} = \frac{\Omega_{\rm m}}{4(1-\Omega_{\rm m})} \frac{\int d\ln a \: H^{-1}}{\int d\ln a \: H^{-1} a^3 y^3} = 1 \label{eq:Lambdacond}
\end{equation}
for the residual cosmological constant from the extended sequestering mechanism to reproduce the value $\Lambda_{\rm obs}$ determined by observations.
Using that $a(t)y(t)$ is symmetric around $t_{\rm turn}=t_{\rm max}/2$, one finds that
\begin{equation}
 \frac{\Lambda_{\rm res}}{\Lambda_{\rm obs}} = \frac{\Omega_{\rm m}}{4(1-\Omega_{\rm m})} \frac{\int dt}{\int dt \; a^3 y^3} = \frac{\Omega_{\rm m}}{8(1-\Omega_{\rm m})} \frac{t_{\rm max}}{\int_0^{t_{\rm turn}} dt \; a^3 y^3} \,.
\end{equation}
Furthermore, $d (a y)/dt=0$ and $d^2 (a y)/dt^2=0$ at $t_{\rm turn}$ such that $a^3y^3|_{t_{\rm turn}}=\Omega_{\rm m}/(1-\Omega_{\rm m})/2$, which follows from Eq.~(\ref{eq:y}).
Thus,
\begin{equation}
 \int_0^{t_{\rm turn}} dt \; a^3 y^3 < \frac{\Omega_{\rm m}}{2(1-\Omega_{\rm m})} t_{\rm turn} \,,
\end{equation}
which implies that
\begin{equation}
 \frac{\Lambda_{\rm res}}{\Lambda_{\rm obs}} > \frac{1}{4} \frac{t_{\rm max}}{ t_{\rm turn}} = \frac{1}{2} \,.
\end{equation}
The longer the evolution remains at $a^3y^3|_{t_{\rm turn}}$, the closer the ratio approximates $\nicefrac{1}{2}$.

One can adopt a more convenient normalisation for the scale factor $a$ to eliminate the dependence of the ratio in Eq.~(\ref{eq:Lambdacond}) on $\Omega_{\rm m}$. 
As long as the cosmological constant is not strictly vanishing, there is always a time when $\rho_{\rm m}=\rho_{\Lambda}=\Lambda$.
With a normalisation of the scale factor at the time of this equality, one therefore finds $H_{\rm eq} \equiv H(a=a_{\rm eq}\equiv1)$ with the corresponding energy density parameters defined at $a_{\rm eq}\equiv1$ simplifying to $\Omega_{\Lambda}=\Omega_{\rm m}=\nicefrac{1}{2}$.
The prefactor in the second expression of Eq.~(\ref{eq:Lambdacond}), for instance, then simplifies to $\nicefrac{1}{4}$.
More generally, solutions for $\Lambda_{\rm res}/\Lambda_{\rm obs}$ for a given $\delta_i$ are universal and can be rescaled from one $\Omega_{\rm m}$ to another by different normalisations of the scale factor $a$ and a corresponding rescaling of $\delta_i$.
We therefore only need to inspect the ratio $\Lambda_{\rm res}/\Lambda_{\rm obs}$ for $a_{\rm eq}\equiv1$.
Fig.~\ref{fig:ratio} illustrates this ratio as a function of $\delta_i$, which is naturally driven to unity for the latest collapsed structures formed in the future of our Universe.
With some fine tuning in $\delta_i$ the value may further drop to approximate $\nicefrac{1}{2}$, but for the $\delta_i$ forming the ultimate collapsed structures, $\Lambda_{\rm res}/\Lambda_{\rm obs}$ can generally be expected to be $\mathcal{O}(1)$.

\subsection{Cosmic acceleration} \label{sec:cosmicacc}

The residual cosmological constant $\Lambda_{\rm res}$ produced by the extended sequestering mechanisms of Sec.~\ref{sec:extendedseq} in the ultimate collapsed structures was found in Sec.~\ref{sec:residualcc} to be consistent with the cosmological constant adopted in the background evolution $\Lambda_{\rm obs}$ with the general limitation that $\Lambda_{\rm res}/\Lambda_{\rm obs}>\nicefrac{1}{2}$.
It could in principle be assumed to be simply fixed by measurement.
However, while this generally implies that the Universe will at some point undergo a late-time accelerated expansion, it does not address the \emph{Why Now?} coincidence problem for why $\Omega_{\Lambda}\sim\Omega_{\rm m}$ today.
An estimation for the relative magnitudes of these energy densities follows from an inspection of the likelihood of our location in the history of the Cosmos.

\subsubsection{A view from the star-formation history} \label{sec:sf}

As a rough guide to finding a bound on the present value of $\Omega_{\Lambda}$, let us first address the coincidence problem by an inspection of the star-formation history of our Universe.
In the presence of a non-vanishing cosmological constant, we can again consider the time $t_{\rm eq}$ of equality between $\rho_{\Lambda}$ and $\rho_{\rm m}$.
Consider a Hubble constant $H_i \equiv H(t_i \ll t_{\rm eq})$ and the baryonic energy density $\rho_{{\rm b},i}\equiv\rho_{\rm b}(t_i \ll t_{\rm eq}) \equiv f_{\rm b}\rho_{\rm m}(t_i \ll t_{\rm eq})\simeq f_{\rm b}\rho_{\rm tot}(t_i \ll t_{\rm eq})$ in the matter-dominated regime.
The baryonic fraction $f_{\rm b}$ and $H_i$ shall be free parameters of the Universe determined by measurement and independent of the presence and value of a cosmological constant.
Note that the extrapolation of $H_{\rm eq}$ from $H_i$ is independent of the value of $\Lambda$, only assuming it to be non-vanishing.
In the following we will adopt the Planck~\cite{Ade:2015xua} values $f_{\rm b}=0.158$ and $H_{\rm eq}=0.0817\:\textrm{Gyr}^{-1}$.

We shall assume the inductive star-formation rate of Ref.~\cite{Lombriser:2017cjy} based on the empirical fit of Ref.~\cite{Madau:2014bja}, where
\begin{eqnarray}
\Gamma_{\rm sf} & \propto & \frac{f_{\rm in} + w f_{\rm out}}{1+w} \,, \nonumber\\
 f_{\rm in} & = & \frac{1}{2}\left(\frac{3}{2}\right)^2 \left( \frac{\rho_{\rm b}}{2\rho_{\Lambda}} \right)^{1/3} f_{\rm out} \,, \\
 f_{\rm out} & = & \frac{4\rho_{\rm b}\rho_{\Lambda}}{\left(\rho_{\rm b}+\rho_{\Lambda}\right)^2} 
\end{eqnarray}
with weight $w(a) = (a_{\rm infl}/a)^n$, $n=7$, and $a_{\rm infl}$ is determined by $(3+\sqrt{7})\rho_{\Lambda}/[2\rho_{\rm b}(a_{\rm infl})]=1$.
The peak of the star-formation history is located at $\dot{\Gamma}_{\rm sf}=0$, or $f_{\rm in}+wf_{\rm out}=1+w$ due to the normalisation to unity of this ratio at the peak.
With normalisation of $a_{\rm eq}\equiv1$ we find that $a_{\rm peak}=0.458$.
For a normalisation at $a_0\equiv1$, therefore
\begin{equation}
 a_{\rm peak} = 0.458 \left(\frac{\Omega_{\rm m}}{1-\Omega_{\rm m}}\right)^{1/3} = 0.458 \left(\frac{1-\Omega_{\Lambda}}{\Omega_{\Lambda}}\right)^{1/3} \,.
\end{equation}

For simplicity, let us first na\"ively assume that the Sun and the Earth were produced when the likelihood for star formation was high, hence, at the peak of star formation.
One can immediately see that for the peak to lie in the past, one must have $\Omega_{\Lambda}>0.088$.
Next, we can ensure that the peak lies sufficiently far in the past to allow for the emergence of intelligent life by today.
We shall assume that once the Earth has formed, it will take $\gtrsim\mathcal{O}(1)$ billion years for this process.
It follows that
\begin{equation}
 \Delta t = \int_{a_{\rm peak}}^{a_0} \frac{da}{a\:H} = \frac{\sqrt{2}}{H_{\rm eq}}\int_{a_{\rm peak}}^{a_0} \frac{da}{a\sqrt{1+a^{-3}}} = \left. \frac{2\sqrt{2}}{3 H_{\rm eq}} \textrm{ArcSinh} \: a^{3/2} \right|_{a_{\rm peak}}^{a_0} \,,
\end{equation}
using the normalisation $a_{\rm eq}\equiv1$ in the second equality.
Applying the constraint $\Delta t \equiv t_0-t_{\rm peak}>10^9\:\textrm{yrs}$ increases the lower bound to $\Omega_{\Lambda}>0.139$.

Now let us infer an upper bound.
For that we shall consider that the Sun remains approximately 10~Gyr in the main sequence before it will turn into a red giant and make Earth inhabitable.
This yields a constraint of $\Omega_{\Lambda}<0.681$, which shows that our initial assumption that the Sun was formed at the peak of $\Gamma_{\rm sf}$ is observationally not viable.
In fact the star-formation peak is measured to lie at $z\approx1.9$ with the formation of the Sun in contrast corresponding to $z\approx0.4$, thus about 6 billion years after the peak position.

Allowing instead the Sun to have formed within $3\sigma$ of all stars produced with $\Gamma_{\rm sf}$ yields the constraint
\begin{equation}
 0.011<\Omega_{\Lambda}<0.985
\end{equation}
for the same calculation, implying that the energy densities of matter and the cosmological constant today should not differ by more than about a factor of $10^2$.

\subsubsection{Uniform prior on location in collapse} \label{sec:uniformprior}

A simpler approach to estimating $\Omega_{\Lambda}$ than by inspection of the star-formation history and that is intrinsically connected to the extended sequestering mechanism is by adopting a uniform prior on the current location in the formation of the ultimate collapsed structures.
Since the evolution of the top hat is not steady in time, it does not indicate to be the correct quantity to characterise this location in.
Instead we shall consider the dimensionless physical top-hat radius $y$ that evolves from unity with a small perturbation $\delta_i$ at early times to zero at the time of collapse.
Placing a uniform prior on $y$ yields $\langle y \rangle = \nicefrac{1}{2}$.
We shall therefore assume that $y=\nicefrac{1}{2}$ at $t_0$, implying that there is as much of the evolution ahead of us as there is in the past.
Requiring $\Lambda_{\rm res}=\Lambda_{\rm obs}$ and $y(a=1)=\nicefrac{1}{2}$ with the normalisation $a_0\equiv1$ determines $\Omega_{\rm m}$ and $\delta_i$ and yields an energy density parameter for the cosmological constant of
\begin{equation}
 \Omega_{\Lambda}=0.697 \,,
\end{equation}
which is in very good agreement with observations~\cite{Ade:2015xua} and hence reproduces the observed current late-time accelerated expansion of our Universe.
Finally, note that the result implies $t_0=t_{\rm turn}/2=t_{\rm max}/4$.

\subsubsection{Observational implications} \label{sec:observations}

The proposed mechanism for the production of $\Lambda_{\rm obs}$ by the sequestering of vacuum energy with a residual cosmological constant $\Lambda_{\rm res}$ in the formation of the ultimate collapsed structures can naturally be expected to produce fluctuations in the $\Lambda_{\rm res}$ of the different collapsing patches around the averaged background value $\Lambda_{\rm obs}$ (Sec.~\ref{sec:residualcc}).
In particular, one finds that an initial top-hat overdensity $\delta_i$ at $a_i=10^{-2}$ that in the local patch is enhanced by $1.2\%$ over the average $\delta_i$ yielding $\Lambda_{\rm res}=\Lambda_{\rm obs}$ recovers the $8\%$ discrepancy between the local measurement of the Hubble constant $H_0=73.24\pm1.74\:\textrm{km}^{-1}\textrm{Mpc}^{-1}$~\cite{Riess:2016jrr} and its cosmological Planck value $H_0=67.81\pm0.92\:\textrm{km}^{-1}\textrm{Mpc}^{-1}$~\cite{Ade:2015xua}.
The significance in this tension is reported at the $3.8\sigma$ level~\cite{Riess:2018byc} and the discrepancy is therefore worth further investigating in the context of a local deviation in $\delta_i$.
This lies beyond the scope of this work and a detailed analysis of the likelihood of such a deviation in the proposed mechanism as well as of further observational consequences is left to future work.

\section{Conclusions} \label{sec:conclusions}

Identifying the physical nature driving the observed late-time accelerated expansion of our Universe remains a difficult task to cosmologists.
The cosmological constant provides a simple explanation for this acceleration and can be thought attributed to vacuum fluctuations.
Quantum theoretical calculations of this vacuum energy contribution, however, are off by several orders of magnitude.
The Standard Model vacuum energy sequestering mechanism has been proposed as a remedy to the old aspect of the cosmological constant problem, that of the non-gravitating vacuum energy, at least for the matter sector.
But the mechanism does not by itself predict the future collapse of the Universe that is required.
Moreover, in the global version of sequestering, the residual cosmological constant produced by the mechanism, while radiatively stable, is generally too small to explain the observed cosmic acceleration.
In contrast, in the local version, an arbitrary constant correction term to the residual can be fixed to match observations.

In this paper the global sequestering mechanism has been extended to account for patchy matter distributions filling the Cosmos that only becomes homogeneous on large scales.
These regions undergo gravitational collapse.
In the future, structure formation ceases, which produces residual cosmological constants from the ultimate collapsed structures that determine an averaged cosmological constant for the cosmological background.
The size of the resulting cosmological constant was found to be naturally consistent with the observed value.
Moreover, the sequestering term appears naturally in the action of the theory from requiring the residual cosmological constant not to be confined to the collapsing patches but to extend to the empty regions left behind, which defines an averaging condition.

Interestingly, small deviations between the residual cosmological constants produced in the different patches suggest that the local Hubble constant may differ from that of a cosmological average in the background.
The analysis of whether the proposed mechanism can naturally alleviate observational tensions between local and cosmological measurements of the Hubble constant is left subject to future work.

\section*{Acknowledgments}

This work was conducted with support by a Swiss National Science Foundation Professorship grant (No.~170547). Please contact the author for access to research materials.

\bibliographystyle{JHEP}
\bibliography{cc}

\end{document}